\documentclass{IEEEtran4PSCC}

\usepackage{cite}

\ifCLASSINFOpdf
   \usepackage[pdftex]{graphicx}
\else
   \usepackage[dvips]{graphicx}
\fi
\usepackage[cmex10]{amsmath}
\usepackage{amssymb,amsfonts}
\usepackage{amsthm}
\usepackage{mathtools}
\usepackage{nicefrac}

\DeclareMathOperator{\diff}{d}

\newcommand{\R}{\mathbb{R}}

\newcommand{\ddt}{\tfrac{\diff}{\diff \!t}}

\usepackage{colortbl}
\usepackage{xcolor}
\usepackage{booktabs} 
\usepackage{multirow} 
\usepackage{multicol}
\usepackage{balance}

\theoremstyle{plain}
\newtheorem{thm}{Theorem}
\newtheorem{assum}[thm]{Assumption}

\makeatletter
\let\old@ps@headings\ps@headings
\let\old@ps@IEEEtitlepagestyle\ps@IEEEtitlepagestyle
\def\psccfooter#1{%
    \def\ps@headings{%
        \old@ps@headings%
        \def\@oddfoot{\strut\hfill#1\hfill\strut}%
        \def\@evenfoot{\strut\hfill#1\hfill\strut}%
    }%
    \def\ps@IEEEtitlepagestyle{%
        \old@ps@IEEEtitlepagestyle%
        \def\@oddfoot{\strut\hfill#1\hfill\strut}%
        \def\@evenfoot{\strut\hfill#1\hfill\strut}%
    }%
    \ps@headings%
}
\makeatother

\psccfooter{%
        \parbox{\textwidth}{\hrulefill \\ \small{23rd Power Systems Computation Conference} \hfill \begin{minipage}{0.2\textwidth}\centering \vspace*{4pt} \includegraphics[scale=0.06]{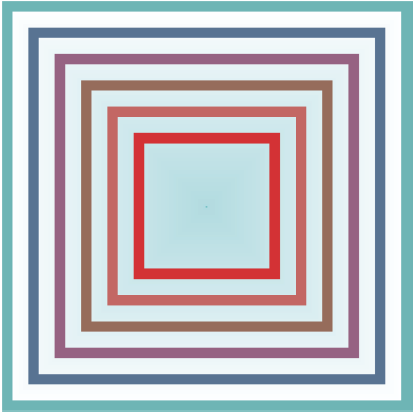}\\\small{PSCC 2024} \end{minipage} \hfill \small{Paris, France --- June 4 -- 7, 2024}}%
}

\begin{document}
%
\title{A Stiffness-Oriented Model Order Reduction Method for Low-Inertia Power Systems}

\author{
\IEEEauthorblockN{Simon Muntwiler$^\star$\IEEEauthorrefmark{2}, Ognjen Stanojev$^\star$\IEEEauthorrefmark{3}, Andrea Zanelli\IEEEauthorrefmark{2}, Gabriela Hug\IEEEauthorrefmark{3}, Melanie N. Zeilinger\IEEEauthorrefmark{2}}
\IEEEauthorblockA{\IEEEauthorrefmark{2}Institute for Dynamic Systems and Control, \IEEEauthorrefmark{3}EEH - Power Systems Laboratory \\
ETH Z\"urich,
Z\"urich, Switzerland\\
\{simonmu, ognjens, zanellia, ghug, mzeilinger\}@ethz.ch}
}


\maketitle

\begin{abstract}
This paper presents a novel model order reduction technique tailored for nonlinear power systems with a large share of inverter-based energy resources. Such systems exhibit an increased level of dynamic stiffness compared to traditional power systems, posing challenges for time-domain simulations and control design. Our approach involves rotation of the coordinate system of a linearized system using a transformation matrix derived from the real Jordan canonical form, leading to mode decoupling. The fast modes are then truncated in the rotated coordinate system to obtain a lower-order model with reduced stiffness. Applying the same transformation to the original nonlinear system results in an approximate separation of slow and fast states, which can be truncated to reduce the stiffness. The resulting reduced-order model demonstrates an accurate time-domain performance, the slow eigenvalues of the linearized system are correctly preserved, and a reduction in the model stiffness is achieved, allowing for accurate integration with increased step size. Our methodology is assessed in detail for a 3-bus system with generation units involving grid-forming/following converters and synchronous machines, where it allows for a computational speed-up of up to $\mathbf{100x}$ compared to the original system. Several standard larger test systems are also considered.
\end{abstract}

\begin{IEEEkeywords}
Low-Inertia Power Systems, Inverter Dynamics, Model Order
Reduction, Stiff Dynamic Systems, Simulation.
\end{IEEEkeywords}

\thanksto{\noindent Submitted to the 23rd Power Systems Computation Conference (PSCC 2024).\\
$^\star$ Simon Muntwiler and Ognjen Stanojev contributed equally to this paper. \\
This research was supported by NCCR Automation, a National Centre of Competence in Research, funded by the Swiss National Science Foundation (grant number 51NF40\_180545).
The work of Simon Muntwiler was supported by the Bosch Research Foundation im Stifterverband.}

\section{Introduction}
The increased penetration of inverter-based energy resources in power systems has significantly changed the dynamic characteristics of the system, as the converter dynamics and their controls operate on a timescale that closely aligns with the network dynamics \cite{Doerfler2023}. Consequently, conventional modeling and simulation approaches relying on the assumption of a timescale separation between line and generation dynamics have become inadequate, necessitating the development of novel power system simulation models. One of the challenges in developing new simulation tools lies in determining the appropriate level of model complexity \cite{Caduff2021}. Recent studies propose detailed modeling of network dynamics, inverter-based resources, and loads to represent all the relevant dynamic phenomena \cite{markovic2021understanding}. Nevertheless, these dynamic models exhibit a significant degree of \textit{stiffness} and a large number of variables, thus becoming intractable even when being solved with state-of-the-art variable time step integration methods. 

Recent literature proposes using model order reduction (MOR) techniques to ensure computational tractability while preserving sufficient modeling detail required to analyze low-inertia systems.
A widely used concept in MOR is to formulate a system in singularly perturbed form, where the slow and fast dynamics are separated.
The model order and stiffness can then be reduced by treating the fast states as algebraic variables~\cite{Sauer1982,Luo2014,Curi2017,Vorobev2018,Dhople2022, Caduff2021,kokotovic1999singular}.
This concept has been employed, for instance, for reducing the order of synchronous machine models \cite{Sauer1982}, eliminating fast-varying dynamics of networks of grid-forming inverters \cite{Dhople2022}, and for constructing a tractable model of low-inertia systems for control design \cite{Curi2017}.
The main challenge, however, is to transform a system into a singularly perturbed form, especially in the case of nonlinear systems.

Various transformation methods to obtain a system in singularly perturbed form exist, such as participation factor analysis (PFA) \cite{Cossart2018,Cossart2022,Grdenic2023}, modal reduction \cite{Price1978,Oliveira1988}, and balanced truncation \cite{Laub1987,Ma1988,Freitas2008,Osipov2018}.
In \cite{Cossart2018,Cossart2022,Grdenic2023}, a separation between fast and slow modes is leveraged based on modal analysis and participation factors \cite{Arriaga1982}. Reduced converter models of single converter units were initially formulated in \cite{Cossart2018} and subsequently extended in \cite{Cossart2022} to converter interconnections, through sequential elimination of the fastest modes based on participation factors. Using a similar methodology, \cite{Grdenic2023}  proposed an approach to identify the transmission lines most critical for system stability and the necessary level of their model complexity. Although participation factors provide supplementary insights, their interpretability is frequently obscured by the co-participation of multiple states in a single mode. Moreover, the aforementioned methods may not consistently preserve the stability characteristics of the original system, as shown in \cite{Caduff2021}.

The approaches based on modal analysis and PFA reviewed above reduce the order of the system by manipulating the original system states and equations, thereby preserving the physical interpretation of variables. However, methods that perform order reduction in a transformed space, such as balanced truncation and modal reduction, generally achieve models of higher quality \cite{ZhuKrylov2019}. The balanced truncation method possesses theoretical advantages over alternative approaches as a global error bound can be established and stability properties are maintained \cite{Antoulas2004}.
However, it requires the calculation of controllability and observability Gramians by solving two dual Lyapunov equations \cite{Freitas2008}, which leads to computational difficulties. While the modal reduction methods \cite{Price1978,Oliveira1988} are also promising, they have previously only been considered in the context of conventional and linear power system models.

This paper proposes a novel MOR technique for nonlinear systems, building upon the traditional modal reduction approaches, directly targeting the reduction of the model stiffness while at the same time preserving the critical eigenvalues\footnote{The eigenvalues of a (stable) linear system with the smallest stability margin (i.e., the eigenvalues with the largest real part) are denoted as critical eigenvalues.} of the linearized system. Therefore, small-signal stability properties of the original system are preserved, and larger integration step sizes are viable without a significant loss in model accuracy. In the proposed approach, the coordinate system of a system linearized around a desired operation point is first rotated using an orthonormal matrix obtained via the real Jordan canonical form, yielding decoupling of the system modes. The fast modes are then truncated in the rotated coordinate system to obtain a lower-order model with reduced stiffness, while the slow modes are entirely preserved. Applying the same transformation to the original nonlinear system results in an approximate separation of slow and fast states, which are transformed into algebraic variables to reduce the model order and stiffness.
In contrast to \cite{Cossart2018,Caduff2021} where only single inverter infinite bus system configurations are considered or \cite{Cossart2022,Curi2017,Dhople2022} where interconnections of only grid-forming converters are considered, we investigate realistic generation portfolios in low-inertia systems, including synchronous machines and grid-forming/following converters. We perform a detailed analysis of the effectiveness of the proposed method and compare it to the state-of-the-art methods from the literature using several standard test transmission networks. 

\section{Preliminaries}
\subsection{Considered DAE Systems}\label{sec:sys}
We consider \emph{stiff} dynamical systems which are described by a nonlinear differential-algebraic equation (DAE) of the form
\begin{subequations} \label{eq:DAE}
\begin{align}
    \ddt {x} &= f(x,u,z), \label{eq:DAE_1}\\
    0 &= g(x,u,z),  \label{eq:DAE_2}
\end{align}
\end{subequations}
where $x\in\mathbb{R}^{n_\mathrm{x}}$ are the differential system states, $u\in\mathbb{R}^{n_{\mathrm{u}}}$ the control inputs, and $z\in\mathbb{R}^{n_{\mathrm{z}}}$ the algebraic variables.
We rely on the following assumption\footnote{Under this assumption, the DAE~\eqref{eq:DAE} is regular, compare~\cite[Prop.~1]{reich1991existence}. Regularity of the DAE~\eqref{eq:DAE} implies existence and uniqueness of a solution for consistent initial values $(x_0,u_0)$ (which due to Assumption~\ref{ass:index-1} uniquely define $z_0$ through~\eqref{eq:DAE_2})~\cite[Thm.~3]{reich1991existence}.} to ensure the system~\eqref{eq:DAE} is numerically solvable~\cite[Sec.~VI.1]{wanner1996solving}.
\begin{assum}\label{ass:index-1}
    The DAE~\eqref{eq:DAE} is index-1, i.e.,
\begin{equation}
    \frac{\partial g(x,u,z)}{\partial z}\Big|_{(x,u,z)=(\bar{x},\bar{u},\bar{z})}
\end{equation}
is full-rank for any $(\bar{x}, \bar{u}, \bar{z})$.
\end{assum}
Linearizing~\eqref{eq:DAE} around a desired operation point ($x_\mathrm{o}$, $u_\mathrm{o}$, $z_\mathrm{o}$) results (locally) in the linear DAE
\begin{subequations} \label{eq:DAE_lin}
        \begin{align}
            \ddt \delta x &= A_{\mathrm{xx}} \delta x + B_{\mathrm{xu}} \delta u + A_{\mathrm{xz}} \delta z + b_{\mathrm{x}}, \label{eq:DAE_lin_1} \\
            0 &= A_{\mathrm{zx}} \delta x + B_{\mathrm{zu}} \delta u + A_{\mathrm{zz}} \delta z + b_{\mathrm{z}}, \label{eq:DAE_lin_2}
        \end{align}
\end{subequations}
where $\delta x = x - x_\mathrm{o}$, $\delta u = u - u_\mathrm{o}$, $\delta z = z - z_\mathrm{o}$, $A_{\mathrm{xx}}$, $B_{\mathrm{xu}}$, and $A_{\mathrm{xz}}$ are the Jacobians of $f$ w.r.t. $x$, $u$, $z$, respectively, $A_{\mathrm{zx}}$, $B_{\mathrm{zu}}$, and $A_{\mathrm{zz}}$ are the Jacobians of $g$ w.r.t. to $x$, $u$, $z$, respectively, and $b_{\mathrm{x}}$ and $b_{\mathrm{z}}$ are the bias terms.
Under Assumption~\ref{ass:index-1}, $A_{\mathrm{zz}}$ is regular and~\eqref{eq:DAE_lin_2} can be rearranged to
\begin{align}
    \delta z = -A_{\mathrm{zz}}^{-1}(A_{\mathrm{zx}}\delta x + B_{\mathrm{zu}} \delta u + b_{\mathrm{z}}).
\end{align}
It follows that~\eqref{eq:DAE_lin} can be equivalently represented by the linear ODE\footnote{Note that under Assumption~\ref{ass:index-1}, application of the implicit function theorem allows us to directly obtain a nonlinear ODE which (locally) represents the original DAE~\eqref{eq:DAE}, see~\cite[Sec.~VI.1]{wanner1996solving}.}
\begin{align} \label{eq:sys_ODE_lin}
    \ddt \delta x =&  \underbrace{(A_{\mathrm{xx}} - A_{\mathrm{xz}} A_{\mathrm{zz}}^{-1}A_{\mathrm{zx}})}_{=\vcentcolon \tilde{A}}\delta x + \underbrace{b_{\mathrm{x}} -A_{\mathrm{xz}} A_{\mathrm{zz}}^{-1}b_{\mathrm{z}}}_{=\vcentcolon\tilde{b}} \nonumber \\
    & +\underbrace{(B_{\mathrm{xu}} - A_{\mathrm{xz}}A_{\mathrm{zz}}^{-1} B_{\mathrm{zu}})}_{=\vcentcolon \tilde{B}}\delta u.
\end{align}
The eigenvalues of $\Tilde{A}$ are denoted as $\lambda_i$ with $i\in\{1,\ldots,n_{\mathrm{x}}\}$ and ordered according to their real part, i.e., $\Re(\lambda_1) \le \Re(\lambda_2)\le\ldots\le\Re(\lambda_{n_\mathrm{x}})$.
The corresponding right and left eigenvectors of $\tilde{A}$ are denoted as $y_i$ and $v_i$, respectively, for $i\in\{1,\ldots, n_{\mathrm{x}}\}$, and defined by
\begin{align}
    \tilde{A} y_i &= y_i \lambda_i,  && y_i \neq 0,\\
    v_i^{\top} \tilde{A} &= \lambda_i v_i^{\top}, && v_i \neq 0.
\end{align}

In Section~\ref{sec:modeling}, we show how to obtain a DAE of the form~\eqref{eq:DAE} for a generic low-inertia power system.
Assumption~\ref{ass:index-1} has been proven to hold for a linear power system DAE~\eqref{eq:DAE_lin} in \cite{Nugroho2022}, whereas a proof for nonlinear DAEs~\eqref{eq:DAE} is still an open problem \cite{Tjorben2016}.

\subsection{Definition of Stiffness}\label{sec:stiffness}
Stiffness in linear ODE systems of the form~\eqref{eq:sys_ODE_lin} is often assessed through the so-called stiffness ratio\cite{lambert1991numerical}, i.e., the ratio 
\begin{equation}\label{eq:stiffness_ratio}
    \rho=\frac{|\Re(\lambda_{1})|}{|\Re(\lambda_{n_{\mathrm{x}}})|}.
\end{equation}
In particular, a system~\eqref{eq:sys_ODE_lin} is denoted as stiff\footnote{As noted in~\cite[Sec.~6.2]{lambert1991numerical}, this definition of stiffness is not entirely satisfactory, since some systems exhibit \emph{no} stiff behavior, even though the stiffness ratio is infinite. Nevertheless, we use this definition here, since the stiffness ratio (of the linearized system) can be directly reduced with our proposed MOR approach. Furthermore, the numerical examples in Section~\ref{sec:num} show that a reduction of the stiffness ratio leads to increased integration accuracy with comparably large step sizes, and thus to stiffness reduction.} if all eigenvalues of $\tilde{A}$ have negative real parts and the stiffness ratio is large~\cite[Sec.~6.2]{lambert1991numerical}.
Stiffness of the original system~\eqref{eq:DAE} is then assessed through the stiffness ratio of the linearized system~\eqref{eq:sys_ODE_lin}.

In more general terms, a system is considered as stiff if implicit integration methods work significantly better compared to explicit ones~\cite[Chap.~IV]{wanner1996solving}.
Consequently, accurate integration of a stiff system~\eqref{eq:DAE} using an explicit method requires extremely small step sizes~\cite[Sec.~6]{lambert1991numerical}.
Thus, simulations of~\eqref{eq:DAE} are computationally expensive and it is in general intractable to use the model for control or estimation purposes.
The objective of this paper is to obtain a stiffness-oriented MOR approach that allows for accurate integration of~\eqref{eq:DAE} using comparably large step sizes.

\section{Model Order Reduction}
A common approach to reduce the order and associated stiffness of a model is the concept of singular perturbation~\cite{Caduff2021,kokotovic1999singular,Sauer1982,Luo2014,Curi2017,Vorobev2018,Dhople2022}.
Assume~\eqref{eq:DAE} can be transformed to the singularly perturbed DAE
\begin{subequations} \label{eq:DAE_singular}
\begin{align}
    \ddt x_{\mathrm{s}} &= f_{\mathrm{s}}(x_{\mathrm{s}},x_{\mathrm{f}},u,z), \label{eq:DAE_singular_1}\\
    \epsilon \cdot \ddt x_{\mathrm{f}} &= f_{\mathrm{f}}(x_{\mathrm{s}},x_{\mathrm{f}},u,z), \label{eq:DAE_singular_2}\\
    0 &= g(x_{\mathrm{s}},x_{\mathrm{f}},u,z), \label{eq:DAE_singular_3}
\end{align}
\end{subequations}
where $x_{\mathrm{s}}$ and $x_{\mathrm{f}}$ represent slow and fast states, respectively, and $\epsilon \in \mathbb{R}_{\ge 0}$ is a small positive scalar.
Additionally, assuming that $\frac{\partial f_{\mathrm{f}}}{\partial x_{\mathrm{f}}}$ is Hurwitz stable, model order reduction can be performed by setting $\epsilon = 0$ leading to the reduced-order DAE
\begin{subequations} \label{eq:DAE_singular_red}
\begin{align} 
    \ddt x_{\mathrm{s}} &= f_{\mathrm{s}}(x_{\mathrm{s}},x_{\mathrm{f}},u,z), \label{eq:DAE_singular_red_1}\\
    0 &= f_{\mathrm{f}}(x_{\mathrm{s}},x_{\mathrm{f}},u,z), \label{eq:DAE_singular_red_2}\\
    0 &= g(x_{\mathrm{s}},x_{\mathrm{f}},u,z), \label{eq:DAE_singular_red_3}
\end{align}
\end{subequations}
with reduced stiffness.
As shown in~\cite[Thm.~6.2]{o1968topics}, the solution to the DAE~\eqref{eq:DAE_singular_red_1}-\eqref{eq:DAE_singular_red_2} recovers the solution to the original ODE~\eqref{eq:DAE_singular_1}-\eqref{eq:DAE_singular_2} for $\epsilon \rightarrow 0$.
However, transforming the original system~\eqref{eq:DAE} to singularly perturbed form~\eqref{eq:DAE_singular} is in general challenging.
In the following, we describe three methods to address this challenge: two classical ones based on PFA (Section~\ref{sec:pfa}) and balanced truncation (Section~\ref{sec:bt}), and our proposed stiffness-oriented MOR approach (Section~\ref{sec:proposed_MOR}).

\subsection{Participation Factor Analysis} \label{sec:pfa}
Participation factor analysis (PFA)~\cite{Arriaga1982} is a tool to analyze the contribution of the states of a linear system~\eqref{eq:sys_ODE_lin} to its modes.
The corresponding participation matrix $P\in\mathbb{R}^{n_{\mathrm{x}}\times n_{\mathrm{x}}}$ associated to $\tilde{A}$ in~\eqref{eq:sys_ODE_lin} can be obtained as
\begin{equation}
    P_{ki} = [v_{i}]_k[y_{i}]_k,
\end{equation}
where $[v]_k$ denotes the $k^{\mathrm{th}}$ entry of vector $v$.
The solution of the autonomous and unbiased system~\eqref{eq:sys_ODE_lin} (i.e.,~\eqref{eq:sys_ODE_lin} with $\delta u = 0$ and $\tilde{b} =0$) can be obtained as
\begin{equation}
    x(t) = \sum_{i=1}^n v_i^{\top}x(0)e^{\lambda_i t}y_i,
\end{equation}
and, if $x(0) = y_i$,
\begin{equation}
    x(t) = v_i^{\top}y_iy_ie^{\lambda_i t} = \sum_{k=1}^n (P_{ki})y_ie^{\lambda_i t}.
\end{equation}
Consequently, the participation factor $P_{ki}$ measures the relative contribution of the $k^{\mathrm{th}}$ state in the time response associated with the $i^{\mathrm{th}}$ mode.
Thereby, the participation matrix $P$ can be used to identify the states that contribute to the fast modes.
Thus,~\eqref{eq:sys_ODE_lin} is transformed to singularly perturbed form by introducing slow and fast states as $x_{\mathrm{s}} = T_{\mathrm{s}}^{\mathrm{pfa}}x$ and $x_{\mathrm{f}} = T_{\mathrm{f}}^{\mathrm{pfa}}x$, respectively, where $T_{\mathrm{s}}^{\mathrm{pfa}}$ is an identity matrix with zeros on its diagonal at entries corresponding to states with large contribution to fast modes and $T_{\mathrm{f}}^{\mathrm{pfa}}=I-T_{\mathrm{s}}^{\mathrm{pfa}}$.
The original system~\eqref{eq:DAE} can be transformed into singularly perturbed form (approximately) with $f_{\mathrm{s}}(x_{\mathrm{s}},x_{\mathrm{f}},u,z) = T_{\mathrm{s}}^{\mathrm{pfa}} f(x,u,z)$ and $f_{\mathrm{f}}(x_{\mathrm{s}},x_{\mathrm{f}},u,z) = T_{\mathrm{f}}^{\mathrm{pfa}} f(x,u,z)$.
While this allows us to reduce the model order, the approach might fail at significantly reducing the stiffness and preserving the original dynamics due to the non-negligible coupling between slow and fast modes, as demonstrated in a numerical example in Section~\ref{sec:num}.
\subsection{Balanced Truncation}\label{sec:bt}
The balancing transformation~\cite{Laub1987,Freitas2008,Osipov2018} can be used to transform a linear system~\eqref{eq:sys_ODE_lin} into a balanced system with states ordered according to their controllability and observability.
Assuming $\tilde{A}$ is asymptotically stable, the reachability and observability Gramians $W_{\mathrm{r}}$ and $W_{\mathrm{o}}$, respectively, can be obtained as the solution of the algebraic Lyapunov equations
\begin{align}
    \tilde{A} W_{\mathrm{r}} + W_{\mathrm{r}}\tilde{A}^{\top} + \tilde{B}\tilde{B}^{\top} &= 0, \\
    \tilde{A} W_{\mathrm{o}} + W_{\mathrm{o}}\tilde{A}^{\top} + \tilde{C}\tilde{C}^{\top} &= 0,
\end{align}
where $\tilde{C}$ is the linearization of an output equation\footnote{Note that in the case with full state measurement considered here, $\tilde{C}$ can be chosen as identity.} $y = h(x,z,u)$.
Denoting the lower triangular Cholesky factors of the Gramians $W_{\mathrm{r}}$ and $W_{\mathrm{o}}$ as $L_{\mathrm{r}}$ and $L_{\mathrm{o}}$, respectively, i.e., $W_{\mathrm{r}}=L_{\mathrm{r}}L_{\mathrm{r}}^{\top}$ and $W_{\mathrm{o}}=L_{\mathrm{o}}L_{\mathrm{o}}^{\top}$, we obtain the singular value decomposition $L_{\mathrm{o}}^{\top}L_{\mathrm{r}} = U \Delta V^{\top}$ and balancing transformation
\begin{align}
    T_{\mathrm{bt}} = L_{\mathrm{r}}V\Delta^{-\nicefrac{1}{2}}.
\end{align}

In~\cite{Ma1988}, the balancing transformation was used to transform a nonlinear system into the form
\begin{subequations}\label{eq:sys_dae_split}
\begin{align}
    \ddt \tilde{x}_{\mathrm{s}} &= T_{\mathrm{s}}T^{-1} f(T\tilde{x},u,z), \label{eq:sys_dae_split_1} \\
    \ddt \tilde{x}_{\mathrm{f}} &= T_{\mathrm{f}}T^{-1} f(T\tilde{x},u,z), \label{eq:sys_dae_split_2} \\
    0 &= g(T\tilde{x},u,z), \label{eq:sys_dae_split_3}
\end{align}
\end{subequations}
with $T=T_{\mathrm{bt}}$ and where $T_{\mathrm{s}}=T_{\mathrm{s}}^{\mathrm{bt}}=\begin{bmatrix} I_{n_{\mathrm{s}}} & 0 \end{bmatrix}$ and $T_{\mathrm{f}}=T_{\mathrm{f}}^{\mathrm{bt}}=\begin{bmatrix} 0 & I_{n_{\mathrm{f}}} \end{bmatrix}$ extract the states with the $n_{\mathrm{s}}\in [0,n_{\mathrm{x}}]$ largest and $n_{\mathrm{f}} = n_{\mathrm{x}}-n_{\mathrm{s}}$ smallest singular values, respectively.
Finally, MOR can be performed\footnote{Note that in contrast, in~\cite{Ma1988,Osipov2018}, MOR was performed by completely ignoring~\eqref{eq:sys_dae_split_2}, while in~\cite{Verhoeven2009} it was approximated using linear dynamics.} by applying the concept of singular perturbation to~\eqref{eq:sys_dae_split} and treating the modes associated with small singular values as algebraic variables, i.e., setting $\ddt \tilde{x}_{\mathrm{f}} = 0$ in~\eqref{eq:sys_dae_split_2}.
Note that, since the balancing transformation orders the states according to their controllability/observability, this MOR approach may fail at reducing the stiffness of the model, as discussed in our numerical example in Section~\ref{sec:num} and, in particular, the eigenvalue spectrum in the lower row of Fig.~\ref{fig:eig_plots}.

\subsection{Proposed Stiffness-oriented Model Order Reduction}\label{sec:proposed_MOR}
To overcome the limitations of classical MOR using PFA and balanced truncation, we propose a stiffness-oriented MOR technique.
The underlying idea is to perform PFA on system~\eqref{eq:sys_ODE_lin} in a rotated coordinate system where slow and fast modes are completely decoupled.
Since $\tilde{A}$ is a real square matrix, we can obtain the real Jordan canonical form\footnote{In case $\tilde{A}$ is diagonalizable and has only real eigenvalues and eigenvectors, we can directly obtain a diagonal system without relying on the real Jordan canonical form.} $\tilde{A}=T_{\mathrm{J}}JT_{\mathrm{J}}^{-1}$ with real transformation matrix $T_{\mathrm{J}}$ obtained as
\begin{align*}
    T_{\mathrm{J}} =& \begin{bmatrix}
        T_1, T_2, \ldots , T_{n_{\mathrm{x}}}
    \end{bmatrix}, \\
    T_i =& \frac{\tilde{T}_i}{\|\tilde{T}_i\|_2} \text{ with } \tilde{T}_i = \begin{cases}
        y_i & \text{if } \lambda_i \text{ is real}, \\
        \Re (y_i) & \text{if } \lambda_i \text{ is complex}, \\
        \Im (y_i) & \text{if } T_{i-1} = \Re (y_{i-1}),
    \end{cases}
\end{align*}
and real block-diagonal Jordan matrix $J=T_{\mathrm{J}}^{-1}\tilde{A}T_{\mathrm{J}}$~\cite[Sec.~3.4]{Horn2012}.
This allows us to transform~\eqref{eq:sys_ODE_lin} to
\begin{align}\label{eq:ODE_bd}
    \ddt \delta \tilde{x} = J\delta\tilde{x} + T_{\mathrm{J}}^{-1}\tilde{B}\delta u + T_{\mathrm{J}}^{-1}\tilde{b},
\end{align}
with state $\delta\tilde{x} = \tilde{x} - \tilde{x}_\mathrm{o} = T_{\mathrm{J}}^{-1}x$ and $\tilde{x}_\mathrm{o} = T_{\mathrm{J}}^{-1}x_\mathrm{o}$.
The resulting system is block-diagonal with each block being associated with a pair of conjugate eigenvalues.
Consequently, the modes in~\eqref{eq:ODE_bd} are completely decoupled and the corresponding PFA matrix $P$ is block-diagonal.
We can extract the states associated with the $n_{\mathrm{s}}\in [0,n_{\mathrm{x}}]$ slowest and $n_{\mathrm{f}} = n_{\mathrm{x}}-n_{\mathrm{s}}$ fastest modes as $\tilde{x}_{\mathrm{s}} = T_{\mathrm{s}}^{\mathrm{J}}  \tilde{x}$ and $\tilde{x}_{\mathrm{f}} = T_{\mathrm{f}}^{\mathrm{J}}\tilde{x}$, respectively, with $T_{\mathrm{s}}^{\mathrm{J}}=\begin{bmatrix} 0 & I_{n_{\mathrm{s}}} \end{bmatrix}$ and $T_{\mathrm{f}}^{\mathrm{J}}=\begin{bmatrix} I_{n_{\mathrm{f}}} & 0 \end{bmatrix}$.
Herein, deciding on the number of slow states $n_{\mathrm{s}}$ is a choice of the user and subject to the trade-off between reduced stiffness and lower accuracy of the resulting reduced model w.r.t. to the original model.
Thus, the system~\eqref{eq:ODE_bd} can easily be brought into singularly perturbed form~\eqref{eq:DAE_singular}.
An (approximate) separation of slow and fast modes in the original (nonlinear) DAE~\eqref{eq:DAE} can be obtained as in~\eqref{eq:sys_dae_split}, with $T=T_{\mathrm{J}}$, $T_{\mathrm{s}}=T_{\mathrm{s}}^{\mathrm{J}}$ and $T_{\mathrm{f}}=T_{\mathrm{f}}^{\mathrm{J}}$.
Note that due to the use of the real Jordan canonical form (compared to the standard (complex) Jordan canonical form), the resulting system of the form~\eqref{eq:sys_dae_split} stays real, and standard integration methods can be used to solve the system.
As in the case of balanced truncation, MOR can be performed by setting $\ddt \tilde{x}_{\mathrm{f}} = 0$ in~\eqref{eq:sys_dae_split_2}.
This results in a model with reduced stiffness which accurately captures the slow dynamics of the original system as can be seen in the numerical examples provided in Section~\ref{sec:num}, as long as the system state stays sufficiently close to the linearization point.
Since the critical eigenvalues of the linearized system are preserved, the proposed MOR approach should in general not affect the stability properties of the nonlinear system.
In contrast to the proposed method, applying PFA to the original system~\eqref{eq:DAE} and MOR based on balanced truncation may both fail at reducing the stiffness of the system.
Note that a similar MOR approach for linear systems was proposed in~\cite{Price1978,Oliveira1988}.

\section{Modeling and Control of Low-Inertia Systems}\label{sec:modeling}
While the MOR technique introduced in Section~\ref{sec:proposed_MOR} is applicable to \emph{any} stiff DAE system of the form~\eqref{eq:DAE}, the focus of this work is on low-inertia power systems.
This section overviews the typical components of these systems, including synchronous machines and grid-forming/following converters. Additionally, dynamics of transmission lines are also included due to their effect on the stability of the system \cite{markovic2021understanding}. The modeling and control are implemented in a Synchronously-rotating Reference Frame (SRF) and a per-unit system.
\subsection{Graph-theoretic Network Modeling \& Line Dynamics}
We consider a transmission power network represented by a connected graph $\mathcal{G}(\mathcal{N},\mathcal{E})$, with $\mathcal{N} \coloneqq \{1,\dots,n\}$ denoting the set of network nodes, and $\mathcal{E} \subseteq \mathcal{N}\times\mathcal{N}$ representing the set of network branches. The node set is partitioned as $\mathcal{N} = \mathcal{N}_\mathrm{sg} \cup \mathcal{N}_\mathrm{gfm} \cup \mathcal{N}_\mathrm{gfl} \cup \mathcal{N}_\mathrm{load}$ to define sets of nodes hosting synchronous generators, grid-forming, grid-following converters, and loads. For every bus $i\in\mathcal{N}$, let $v_i\in\R^2$ denote the associated voltage vector (comprised of a $d$ and a $q$ component). For each line $(i,j)\in\mathcal{E}$, let $r_{ij}\in\R_{\geq 0}$ and $l_{ij}\in\R_{\geq 0}$ denote its respective resistance and inductance values, and $i_{ij}\in\R^2$ represent the corresponding branch current. The transmission grid is modeled by the following differential equation, written in a dq-frame rotating at frequency $\omega_g$ as:
\begin{equation}
    \ddt{i}_{ij}=\frac{\omega_b}{\ell_{ij}}(v_{i}-v_{j})-\left(\frac{r_{ij}}{\ell_{ij}}\omega_b+\mathcal{J}\omega_b \omega_g\right)i_{ij}, \forall (i,j)\in\mathcal{E} \label{eq:lineCurrent}
\end{equation}
with $w_b$ representing the base frequency, and $\mathcal{J}$ denoting the $2\times 2$, 90-degree rotation matrix.   
\subsection{Power Converter Models} \label{subsec:sys_level}
\begin{figure}[!t]
    \centering
    \includegraphics[scale=0.9]{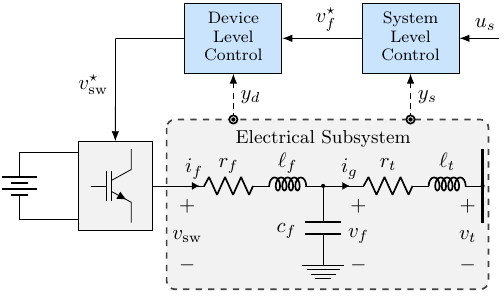}
    \caption{General converter configuration scheme.}
    \label{fig:converter_diag}
\end{figure}

The considered converter model (presented in Fig.~\ref{fig:converter_diag}) consists of a two-level control structure, a switching unit with a constant DC-input voltage, and an AC subsystem, including an RLC filter $(r_f,c_f,\ell_f)$ and a transformer equivalent $(r_t,\ell_t)$.  In the considered control structure, an outer \textit{system-level} control layer provides a reference for the converter's output voltage $v_f^\star$ that is subsequently tracked by a \textit{device-level} controller through direct adjustment of the switching voltage $v_\mathrm{sw}\coloneqq v_\mathrm{sw}^\star$.

\subsubsection{System-level Control} The input measurements of the system-level controller are the filter voltage $v_f \in \R^2$ and the converter current injection into the system $i_g \in \R^2$, collected in $y_s=(v_f,i_g)\in\R^4$. Using these measurements, we calculate the instantaneous active $p_c\coloneqq v^\mathsf{T}_f i_g$ and reactive $q_c \coloneqq v^\mathsf{T}_f \mathcal{J}^\mathsf{T} i_g$ power. The voltage phasor reference is generated by adjusting the predefined active and reactive setpoints $(p^\star_c,q^\star_c)\in\R^2$ according to a measured power imbalance:
\begin{subequations} \label{eq:powersync}
\begin{align} 
	\omega_c &= \omega_s + \tilde{\omega}_c,\quad \ddt{\tilde{\omega}}_c = -\omega_z\tilde{\omega}_c + R_c^p\omega_z(p_c^\star-p_c),\\
	v_c^d &= V_c^\star + \tilde{v}_c,\quad \ddt{\tilde{v}}_c = -\omega_z\tilde{v}_c + R_c^q\omega_z(q_c^\star-q_c),
\end{align}
\end{subequations}
with $R_c^p\in\R_{>0}$ and $R_c^q\in\R_{\geq 0}$ are the droop gains, $\omega_s$ the synchronization frequency, $\omega_z\in\R_{>0}$ denoting the low-pass filter cut-off frequency, $\ddt{\theta}_c = \omega_b\omega_c$ and $v_c^q=0$. As an additional degree of freedom for stabilization and disturbance rejection, virtual impedance $(r_v,\ell_v)\in\R^2_{\geq0}$ of the following form $v_f^\star = v_c - r_v i_g  - \mathcal{J}\omega_c \ell_v  i_g$ is commonly implemented.

\textit{Grid-Following Mode:} A key component of grid-following converters is a synchronization device, commonly a Phase Locked Loop (PLL), which estimates the phase angle $\theta_s \in [-\pi,\pi)$ of the voltage $v_f$ and the grid frequency $\omega_s\in\R_{>0}$:
\begin{equation}\label{eq:theta_pll}
    \omega_s = \omega_g + K_P^s v_f^q + K_I^s \varepsilon,\quad \ddt{\varepsilon} = v_f^q,\quad
    \ddt{\theta}_s = \omega_b\omega_s,
\end{equation}
where $(K_P^s,K_I^s)$ are the proportional and integral control gains of the synchronization unit, and $\varepsilon\in\R$ is the integrator state. Therefore, the synchronization frequency $\omega_s$ in \eqref{eq:powersync} is for grid-following converters defined by the PLL dynamics \eqref{eq:theta_pll}.

\textit{Grid-Forming Mode:} In contrast to the grid-following mode of operation, the synchronization unit is not needed for grid-forming control as $\omega_s=\omega_c^\star$ is set to a constant reference. In this way, these units self-synchronize to the power grid by determining the converter frequency and voltage depending on the output power deviations and without the need for PLL.

\subsubsection{Device-Level Control}
Given a voltage reference $v^\star_f \in \R^2$ in $dq$-coordinates defined by $(\theta_c,\omega_c)$, the device-level control is described by a cascade of voltage and current controllers computing a switching voltage reference $v_\mathrm{sw}^\star\in\R^2$:
\hspace{-5em}
\begin{subequations} \label{eq:srf_vi}
\begin{align} 
    \ddt{\xi} &=  v^\star_f - v_f, \\
    i_f^\star &= K_P^v (v^\star_f - v_f) + K_I^v \xi + K_F^v i_g + \mathcal{J}\omega_c c_f v_f, \label{eq:srf_v_b}
    \end{align}
providing an internal current reference $i_f^\star$ followed by
\begin{align} 
    \ddt{\gamma} &= i_f^\star - i_f,\\
    v_\mathrm{sw}^\star &= K_P^i (i_f^\star - i_f) + K_I^i \gamma + K_F^i v_f + \mathcal{J}\omega_c \ell_f i_f, \label{eq:srf_i_b}
    \end{align}
\end{subequations}
where $(K_P^v,K_P^i)\in\R^2_{>0}$, $(K_I^v,K_I^i)\in\R^2_{\geq0}$ and $(K_F^v,K_F^i)\in\mathbb{Z}_{\{0,1\}}^2$ are the respective proportional, integral, and feed-forward gains, $\xi\in\R^2$ and $\gamma\in\R^2$ represent the integrator states, and $v$ and $i$ indicate the voltage and current controllers.
\subsection{Synchronous Generator Model}
We consider a detailed 14th-order model for synchronous generators,  including transients in the rotor circuit, swing equation dynamics, a round rotor model equipped with a prime mover and a TGOV1 governor, an AVR based on a simplified excitation system SEXS together with a PSS1A power system stabilizer. Similarly to the converters, the synchronous generator is interfaced to the grid through a transformer and modeled in an SRF. ENTSO-E reference \cite{entsoeGen} provides more information on the control configuration and tuning parameters, and \cite{Kundur1994} is suggested for a deeper understanding of the related theory.

\subsection{Complete Model}
Collecting the algebraic and differential states of the considered units \eqref{eq:powersync}-\eqref{eq:srf_vi} and network equations \eqref{eq:lineCurrent} in vectors $x$ and $z$, as well as their respective inputs in $u$, we obtain a system of the form given in \eqref{eq:DAE}. In the considered system, the fastest dynamics correspond to network dynamics and device-level controllers of converter units, with time constants ranging from 1 to 30 ms. On the other hand, the slowest dynamics are related to the mechanical subsystem of synchronous machines with time constants of approx. 10s. The large spread between the timescales of converter-based and synchronous generation controls and components gives rise to model stiffness, as confirmed by the case studies presented in the following.

\section{Numerical Results}\label{sec:num}
We present numerical results for a simple 3-bus and various large-scale systems, all involving inverter-based resources.
\begin{figure}[!t]
    \centering
    \vspace{-0.6cm}
    \includegraphics[width=\columnwidth]{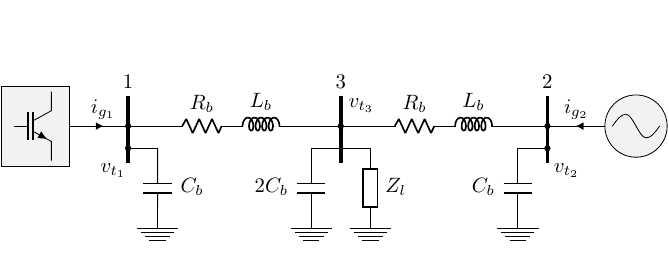}
    \caption{Single line diagram of the considered 3-bus system.}
    \label{fig:3bus_diag}
\end{figure}
\subsection{Reducing the Model Order of a 3-bus Low-Inertia System}\label{sec:num_3bus}
To numerically evaluate the proposed MOR approach, we first consider the simple 3-bus network ($\mathcal{N}=\{1,2,3\}, \mathcal{E}=\{(1,3),(2,3)\}$) depicted in Fig.~\ref{fig:3bus_diag}, where two generators are supplying a constant impedance load via long-distance transmission lines. To test the approach on different combinations of generators with diverse characteristics in low-inertia systems, we investigate interactions of all three unit types introduced in Sec.~\ref{sec:modeling}: 
\begin{itemize}
    \item Scenario~1: A grid-forming converter and a synchronous generator ($\mathcal{N}_\mathrm{gfm}=\{1\},\mathcal{N}_\mathrm{gfl}=\emptyset,\mathcal{N}_\mathrm{sg}=\{2\}$).
    \item Scenario~2: A grid-forming and a grid-following converter ($\mathcal{N}_\mathrm{gfm}=\{1\},\mathcal{N}_\mathrm{gfl}=\{2\},\mathcal{N}_\mathrm{sg}=\emptyset$).
    \item Scenario~3: A synchronous generator and a grid-following converter ($\mathcal{N}_\mathrm{gfm}=\emptyset,\mathcal{N}_\mathrm{gfl}=\{1\},\mathcal{N}_\mathrm{sg}=\{2\}$).
\end{itemize}

In all three scenarios, the load is set to consume 1 p.u. of active power at nominal voltage, the line parameters are $r_{ij}=0.0146\,\mathrm{p.u.},\ell_{ij}=0.146\,\mathrm{p.u.},\forall(i,j)\in\mathcal{E}$, and the generators are set to share the load equally. Converter parameters are adopted from~\cite[Table~1]{markovic2021understanding} and synchronous machine parameters from~\cite[Table~2]{markovic2021understanding}.
The considered test system configurations are implemented in \textsc{MATLAB} using CasADi~\cite{andersson2019casadi} for symbolic modeling. We mainly employ Acados \cite{verschueren2022acados} for numerical integration and occasionally resort to \textsc{MATLAB}'s \texttt{ode15s} variable time-step solver for benchmarking purposes.

\begin{table}[!b]
\vspace{-0.35cm}
\renewcommand{\arraystretch}{1.2}
\caption{Participation Factor Analysis of the 3-bus System.}
\label{tab:partFac}
\noindent
\centering
    \begin{minipage}{\linewidth}
    \renewcommand\footnoterule{\vspace*{-5pt}}
    \begin{center}
    \scalebox{1}{%
        \begin{tabular}{ c || c | c | c | c }
            \toprule
            \textbf{Configuration} & $\boldsymbol{\rho}$ & \textbf{Fastest Modes} & \textbf{States} & \textbf{PF} \\ 
            \cline{1-5}
            \multirow{4}{*}{Scenario 1} & \multirow{4}{*}{5.4E4} & $-4565\pm\mathrm{j}478$ & $i_g,i_{13}$ & $0.75$\\
             & & $-2252\pm\mathrm{j}689$ &  $v_f,i_{23}$ & $0.88$\\
             & & $-1271\pm\mathrm{j}3785$ & $v_f$ & $0.67$\\
             & & $-1181\pm\mathrm{j}3787$ & $v_f,i_g$ & $0.86$\\
            \arrayrulecolor{black!30}\hline
            \multirow{7}{*}{Scenario 2} & \multirow{7}{*}{3.0E4} & $-4678\pm\mathrm{j}356$ &  $i_g,i_{13},i_{23}$ & $0.99$\\
             & & $-3641\pm\mathrm{j}231$ & $i_g$ & $0.6$\\
             & & $-1953\pm\mathrm{j}3995$ & $v_f$ & $0.66$\\
             & & $-1887\pm\mathrm{j}3720$ & $v_f$ & $0.64$\\
             & & $-841\pm\mathrm{j}0$ & $v_f,i_{23}$ & $0.78$\\
             & & $-799\pm\mathrm{j}0$ & $v_f$ & $0.64$\\
             & & $-698\pm\mathrm{j}4131$ & $v_f,\tilde{\omega}_c$ & $0.66$\\
             & & $-667\pm\mathrm{j}3954$ & $v_f,\tilde{\omega}_c$ & $0.64$\\
            \arrayrulecolor{black!30}\hline
            \multirow{4}{*}{Scenario 3} & \multirow{4}{*}{4.4E4} & $-4554\pm\mathrm{j}451$ & $i_g,i_{13},i_{23}$ & $0.98$\\
             & & $-2286\pm\mathrm{j}456$ &  $v_f,i_{13}$ & $0.85$\\
             & & $-1293\pm\mathrm{j}3946$ & $v_f,i_g$ & $0.85$\\
             & & $-1160\pm\mathrm{j}3698$ & $v_f,i_g,i_{23}$ & $0.95$\\
            \arrayrulecolor{black}\bottomrule
        \end{tabular}
    }
        \end{center}
    \end{minipage}
\end{table}

The proposed approach is evaluated against a standard method in the literature \cite{Caduff2021,Cossart2018,Cossart2022,Grdenic2023}, where the states to be removed are selected based on their participation in the fastest modes, as discussed in Sec.~\ref{sec:pfa}. Table~\ref{tab:partFac} showcases the stiffness ratio and the fastest modes for each considered scenario and details the states with the highest participation factors in these modes. As can be seen from Table~\ref{tab:partFac}, all three scenarios are characterized by a high stiffness ratio. To perform PFA-based model reduction, we residualize the states indicated in the table for each respective scenario. As expected, the states to be removed are related to the electrical variables of the converter and the network. On the other hand, the fast modes indicated in Table~\ref{tab:partFac} are removed in the proposed stiffness-oriented MOR after rotating the original state space, as discussed in Sec.~\ref{sec:proposed_MOR}. For both considered MOR approaches, in Scenario 1 and 3, the number of eliminated modes is 8, while in Scenario 2 it is~14.

\begin{figure*}[!t]
    \centering
    \includegraphics[width=\textwidth]{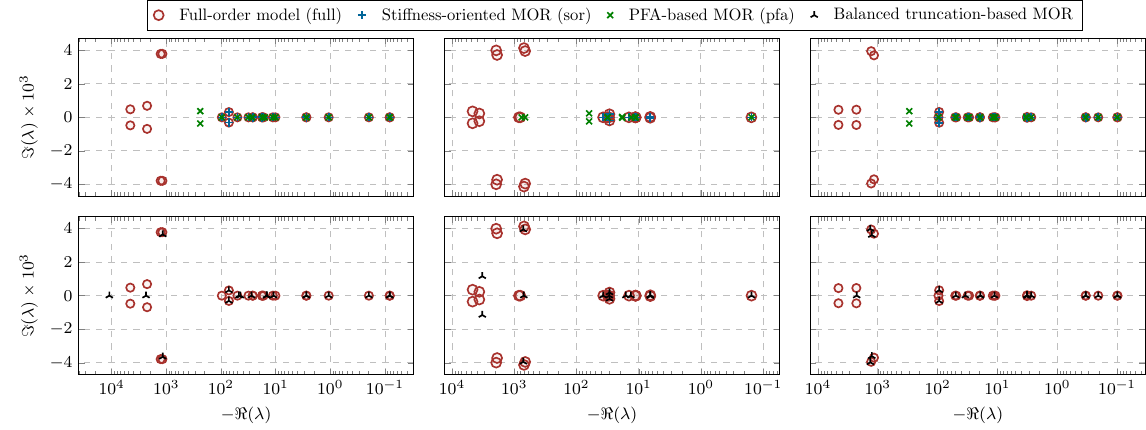}
    \caption{Eigenvalue spectrum of the linearized full and reduced models for Scenario 1 (left column), Scenario 2 (middle column), and Scenario 3 (right column). The upper row shows the eigenvalue spectrum of the reduced models obtained using the proposed stiffness-oriented MOR (Section~\ref{sec:proposed_MOR}) and PFA (Section~\ref{sec:pfa}) while the lower row shows the one of the models obtained using balanced truncation (Section~\ref{sec:bt}).}
    \label{fig:eig_plots}
    \vspace{-0.35cm}
\end{figure*}
\subsection{Modal Analysis}\label{sec:num_modal}
In the upper row of Fig.~\ref{fig:eig_plots}, we present eigenvalue spectrum plots for the linearized original system model ($\mathrm{full}$), the model reduced through PFA ($\mathrm{pfa}$), and the model reduced using the proposed stiffness-oriented MOR ($\mathrm{sor}$) across the three considered scenarios. Our findings indicate that the proposed method effectively preserves the slowest modes while significantly reducing stiffness. More precisely, the new stiffness ratios of the three scenarios are respectively $\rho_1^\mathrm{sor}=1.1\mathrm{E}3,\rho_2^\mathrm{sor}=2.4\mathrm{E}2,\rho_3^\mathrm{sor}=9.3\mathrm{E}2$. On the other hand, the PFA-based approach clearly modifies certain modes. The stiffness ratios of the PFA reduced models are $\rho_1^\mathrm{pfa}=5.2\mathrm{E}3,\rho_2^\mathrm{pfa}=5\mathrm{E}3,\rho_3^\mathrm{pfa}=4.5\mathrm{E}3$. Finally, it is worth noting that for lower-order models PFA-based reduction results in significant modifications of the critical eigenvalues, as also found in \cite{Caduff2021}.
In addition, the lower row in Fig.~\ref{fig:eig_plots} shows the eigenvalue spectrum of reduced-order models obtained using balanced truncation. It can be seen that the stiffness is significantly larger compared to $\mathrm{pfa}$, and thus, we do not consider this MOR method in further comparisons.

\subsection{Accuracy of the Reduced-order Models}\label{sec:num_accuracy}
\begin{figure}[!b]
    \vspace{-0.4cm}
    \centering
    \includegraphics{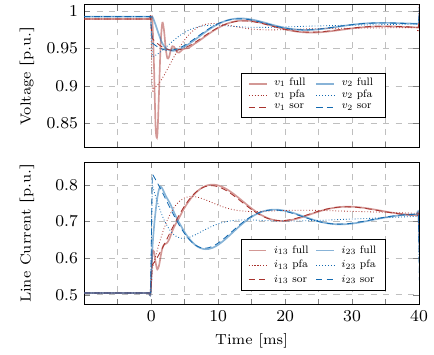}
    \includegraphics{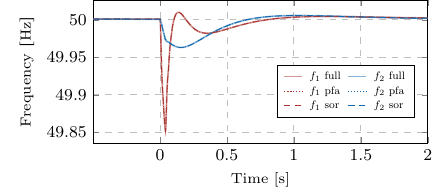}
    \caption{Time domain solution of distinct slow and fast states for Scenario~1.}
    \label{fig:time_domain_plot}
\end{figure}
In Fig.~\ref{fig:time_domain_plot}, we show the time domain evolution of selected fast modes (voltages and line currents) and slow modes (frequencies) of the considered 3-bus system in Scenario~1 after a 30\% load drop at time $t=0$.
The integration is performed using Acados with implicit Runge-Kutta (Gauss-Legendre of order~$4$) and a (small) step size of $h=2.5E-4$ for all three considered models, i.e., $\mathrm{full}$, $\mathrm{sor}$, and $\mathrm{pfa}$.
It can be seen that all models capture the slow dynamics very accurately.
However, the PFA-based approach leads to large integration errors for the fast states, while the model based on the proposed reduction technique is significantly more accurate. Interestingly, the very fast voltage dynamics appear to be averaged immediately following the load loss. 

\begin{figure*}[!t]
    \centering
    \includegraphics[width=\textwidth]{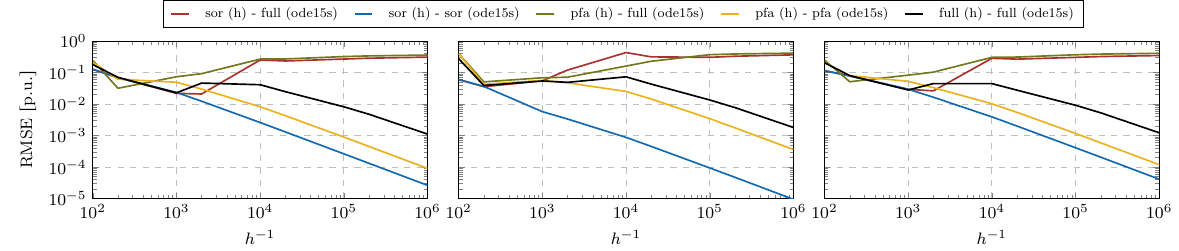}
    \caption{RMSE vs. inverse of step size ($h^{-1}$) for different combinations of models (the reduced model obtained using the proposed stiffness-oriented MOR ($\mathrm{sor}$), the reduced model using PFA ($\mathrm{pfa}$) and the original full model ($\mathrm{full}$) for Scenario 1 (left), Scenario 2 (middle), and Scenario 3 (right)).}
    \label{fig:accuracy_plots}
    \vspace{-0.35cm}
\end{figure*}

\begin{figure}[!b]
    \vspace{-0.35cm}
    \centering
    \includegraphics[width=0.9\columnwidth]{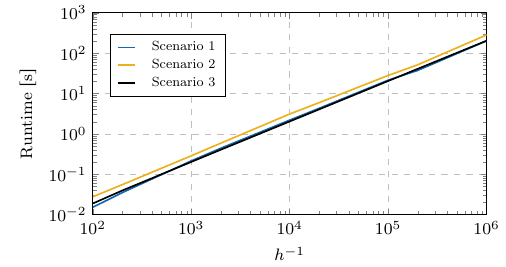}
    \caption{Total runtime vs. inverse of step size ($h^{-1}$) for the 3-bus system model reduced with the proposed MOR ($\mathrm{sor}$) for all three scenarios.}
    \label{fig:time_plots}
    \vspace{-0.35cm}
\end{figure}

To assess the model accuracy in more detail across all considered scenarios and to quantify the computational benefit resulting from the presented reduced-order models, we evaluate the discrepancy between two models ($\mathrm{mod1}$ and $\mathrm{mod2}$) within the time span $t\in[0,T_{\mathrm{end}}]$ with end time $T_{\mathrm{end}}=0.1\mathrm{s}$ through the normalized root mean square error (RMSE), calculated as
\begin{align*}
    \mathrm{RMSE}(h) = \sqrt{\frac{h}{T_{\mathrm{end}}}\sum_{i=1}^{T_{\mathrm{end}}\cdot h^{-1}}\frac{\|x_{\mathrm{mod}1}^{h}(i \cdot h) - x_{\mathrm{mod}2}^{\mathrm{ode15s}}(i \cdot h)\|_2^2}{\|x_{\mathrm{mod}2}^{\mathrm{ode15s}}(i \cdot h)\|_2^2}},
\end{align*}
where $x_{\mathrm{mod}1}^{h}(t)$ is obtained by integration of $\mathrm{mod1}$ with fixed time step $h$ and $x_{\mathrm{mod}2}^{\mathrm{ode15s}}(t)$ by integration of $\mathrm{mod}2$ using the computationally expensive but very accurate solver \texttt{ode15s}.

In Fig.~\ref{fig:accuracy_plots}, we show the RMSE w.r.t. to the inverse step size $h^{-1}$ of the integrator for all three scenarios and different model combinations.
Examining the blue, yellow, and black curves which show the RMSE error of all models ($\mathrm{sor/pfa/full}$) when integrated with the fixed step size $h$ and \texttt{ode15s}, it becomes clear that $\mathrm{sor}$ results in the same RMSE with up to $10\mathrm{x}/100\mathrm{x}$ larger step size compared to $\mathrm{pfa/full}$.
In Fig.~\ref{fig:time_plots}, we additionally show the runtime\footnote{On our system with 12-core Intel i7 processor and 32 GB of memory we obtained an average computational time of $2.3\,$ms per integration step for the $\mathrm{sor}$ model of the considered 3-bus system over all three scenarios.} to simulate until $T_{\mathrm{end}}$ for all three scenarios and models obtained with the proposed MOR approach. It can be seen that an increase of the step size by factor $10\mathrm{x}/100\mathrm{x}$ leads to a reduction of the integration time by factor $10\mathrm{x}/100\mathrm{x}$, respectively.
This result highlights the computational benefit of the proposed approach which is especially promising for step size control~\cite[Chap.~IV]{wanner1996solving}. Step size control is applied in variable time step solvers that reduce the step size in an adaptive manner during integration whenever the error relative to the exact solution is above a certain threshold.

Additionally, by examining the red and green curves which show the RMSE of $\mathrm{sor/pfa}$ integrated with step size $h$ and the full-order model integrated with \texttt{ode15s}, we can see that the proposed approach captures the model dynamics more accurately.
The exceptions, mainly in Scenario~2, where $\mathrm{pfa}$ is more accurate for specific step sizes, arise as $\mathrm{pfa}$ captures very fast changes in states better (e.g., voltage dips), as can also be observed in Fig.~\ref{fig:time_domain_plot} around $t=0$. Thereby, Scenario~2 highlights the trade-off between a reduction in computational cost and resulting accuracy compared to the full model.

\begin{table}[!b]
\vspace{-0.5cm}
\renewcommand{\arraystretch}{1}
\caption{Performance of MOR on larger test cases.}
\label{tab:large_scale_comparison}
\noindent
\centering
    \begin{minipage}{\linewidth} 
    \begin{center}
\scalebox{1}{%
    {\setlength{\extrarowheight}{.15em}\tabcolsep=4pt
    \begin{tabular}{c||c |c c c | c c}
    \toprule
    \multirow{2}{*}{Test Case} & \multirow{2}{*}{\begin{tabular}{c} \textbf{\# Fast} \\ \textbf{Modes}  \end{tabular}} & \multicolumn{3}{c|}{$\boldsymbol{\rho}$} & \multicolumn{2}{c}{\textbf{RMSE}}  \\
            \cline{3-7}
             & & orig & sor & pfa & sor & pfa \\
    \hline
    \begin{tabular}{c} 9-bus IEEE  \end{tabular} & 18 & 4E5 & 9E2 & 1E4 & 0.1097 & 0.1439\\
    \begin{tabular}{c} 3AreaKundur \end{tabular} & 40 & 2E11 & 8E2 & 5E3 & 0.5292 & 0.6303 \\
    \begin{tabular}{c} 39-bus IEEE  \end{tabular} & 104 & 6E11 & 2E2 & 6E2 & 0.2524 & 0.3172\\
    \arrayrulecolor{black}\bottomrule
    \end{tabular}}
    }
    \end{center}
    \end{minipage}
\end{table}
\subsection{Reducing the Model of Larger-scale Low-Inertia Systems}\label{sec:num_large_scale}
After acquiring a fundamental understanding of the considered MOR techniques in a simplified low-inertia test environment, we expand our analysis to larger test cases. Specifically, we examine a 9-bus system, which includes one generator of each type, totaling three generators, a 3-area test system with two units of the same type in each area, and a modified IEEE 39-bus system with three grid-forming units and seven synchronous machines. The results in Table~\ref{tab:large_scale_comparison} demonstrate that the benefits of the proposed MOR scheme extend to larger test cases. In particular, the proposed approach consistently outperforms the PFA approach in both the model accuracy and the reduction of the model stiffness, indicating the possibility of using larger integration steps, which is of particular importance in studies of large-scale systems.
For such large-scale systems, deciding on the states to be removed using PFA is non-trivial, which is overcome by the proposed approach. The RMSE is computed following the same procedure as in the previous subsection, with the integration step of $h=1\mathrm{E}-5$. The same number of states are reduced for both approaches.

\section{Conclusions}
In this paper, we proposed a stiffness-oriented model order reduction (MOR) approach which was shown to effectively reduce the stiffness of low inertia power system models.
Consequently, an up to $100\mathrm{x}$ speed-up of the integration compared to the original (stiff) model allows us to drastically reduce the computational cost while maintaining accurate integration.
The resulting reduced-order modeling is especially promising for optimization-based estimation and control, which we intend to investigate in future work.
Analyzing the effect of operation point changes on the quality of the reduced model is an interesting avenue for future research.

\bibliographystyle{IEEEtran}    
\bibliography{mybib}            
% ...
\appendices
\end{document}